\documentclass[sigconf]{acmart}

\usepackage{amsmath}
\usepackage{booktabs}
\usepackage{graphicx}

\AtBeginDocument{
  }

\copyrightyear{2026}
\acmYear{2026}
\setcopyright{cc}
\setcctype{by}
\acmConference[CIKM '26]{Proceedings of the 35th ACM International Conference on Information and Knowledge Management}{November 07--11, 2026}{Rome, Italy}
\acmBooktitle{Proceedings of the 35th ACM International Conference on Information and Knowledge Management (CIKM '26), November 07--11, 2026, Rome, Italy}
\acmDOI{10.1145/3799682.3840113}
\acmISBN{979-8-4007-2539-5/2026/11}

\begin{document}

\title{Task-Blind No MORE: Multi-Task Information Flow in Unified Ranking Backbones}

\author{Yuchen Wang}
\orcid{0009-0008-2208-2808}
\authornote{Corresponding author.}
\email{wang.yuchen@hellogroup.com}
\affiliation{
  \institution{Hello Group}
  \city{Beijing}
  \country{China}
}

\author{Feng Niu}
\orcid{0009-0001-9911-4029}
\email{niufeng25@mail.ustc.edu.cn}
\affiliation{
  \institution{University of Science and Technology of China}
  \city{Hefei}
  \country{China}
}

\author{Qing Tan}
\orcid{0009-0008-0552-2840}
\email{tanqing24@mails.ucas.ac.cn}
\affiliation{
  \institution{Institute of Software, Chinese Academy of Sciences}
  \city{Beijing}
  \country{China}
}
\affiliation{
  \institution{University of Chinese Academy of Sciences}
  \city{Beijing}
  \country{China}
}

\author{Junting Lu}
\orcid{0009-0008-7999-8112}
\email{lujunting@bistu.edu.cn}
\affiliation{
  \institution{Beijing Information Science and Technology University}
  \city{Beijing}
  \country{China}
}

\author{Baoxin Wu}
\orcid{0009-0003-6736-4044}
\email{wu.baoxin@hellogroup.com}
\affiliation{
  \institution{Hello Group}
  \city{Beijing}
  \country{China}
}

\author{Jun Gao}
\orcid{0009-0002-0372-0268}
\authornotemark[1]
\email{gao.jun@hellogroup.com}
\affiliation{
  \institution{Hello Group}
  \city{Beijing}
  \country{China}
}

\renewcommand{\shortauthors}{Yuchen Wang et al.}

\begin{abstract}
Industrial ranking models for recommendation have scaled feature interaction and sequence modeling separately; recent architectures such as HyFormer and MixFormer unify both in a stackable backbone. Real-world recommender systems, however, nearly always require multi-task learning, yet existing unified architectures confine multi-task modeling to shallow post-backbone towers, leaving the backbone without task-aware information flow.
We propose MORE (Multi-task cO-evolving Ranking modEl), which embeds multi-task information flow inside the backbone, enabling task-specific signals to co-evolve with sequence and feature representations at every layer rather than in a post-hoc fusion. It introduces Anchor Tokens that persist across backbone layers: Shared Anchors encode cross-task commonalities, while Private Anchors capture task-specific priors. In each block, Anchor Tokens (1) read task-conditioned signals from behavior sequences, (2) mix with non-sequential features under a task-boundary mask, and (3) refine per-task representations through independent branches; as blocks stack, each task obtains a differentiated representation refined through all backbone layers.

Experiments on large-scale industrial datasets show that MORE consistently outperforms baselines across all tasks under comparable parameter and FLOPs budgets, and scales well with model size. Online A/B tests on Momo, a leading Chinese social discovery platform with tens of millions of monthly active users, yield 3\% improvement in usage duration, 3.6\% in interaction rate, and 2\% in deep-chat rate. MORE is deployed in production with request-level shared computation reducing scoring latency by about 30\%.
\end{abstract}



\maketitle

\section{Introduction}

Industrial ranking models in recommender systems have evolved along two parallel directions. One direction focuses on feature interaction scaling. Representative methods such as Wukong~\cite{zhang2024wukong} and RankMixer~\cite{rankmixer2025} scale the interaction backbone to improve high-order feature crossing. The other direction focuses on behavior sequence scaling. Representative methods such as HSTU~\cite{zhai2024hstu} and LONGER~\cite{longer2025} allow models to process much longer behavior histories via scalable generative architectures. These two directions remain architecturally separate, preventing bidirectional interaction between sequential and non-sequential features. Recent unified architectures therefore place sequence modeling and feature interaction into the same stackable backbone, as in OneTrans~\cite{onetrans2025}, HyFormer~\cite{hyformer2026}, and MixFormer~\cite{mixformer2026}, advancing both offline accuracy and online scalability.

Yet none of these advances addresses another central requirement in industrial ranking: multi-task learning. Industrial ranking scenarios are almost always multi-task, requiring a single model to jointly serve tasks with heterogeneous behavioral semantics. Although MMoE~\cite{ma2018mmoe}, PLE~\cite{tang2020ple}, and related methods have advanced multi-task modeling considerably, ranking backbones still isolate task information in post-backbone prediction heads. The backbone is therefore task-agnostic. Three limitations follow.
\begin{itemize}
\item \textbf{Task-blind sequence reading and feature interaction}: Task signals enter exclusively at prediction heads and never shape how the backbone organizes information internally. Sequence reading cannot condition on which task it serves, nor can feature interaction adjust its aggregation to task-specific preferences. For instance, click-through rate depends on exposure recency and visual salience, whereas comment rate relies on content depth and topic relevance; a task-agnostic shared query conflates both needs into one undifferentiated reading of the same behavior history.
\item \textbf{Structural bottleneck before task differentiation}: Because all tasks share the same backbone output, task-specific feature preferences are compressed into a single semantic space before any task differentiation begins. Gate and expert mechanisms in post-backbone towers (e.g., MMoE, PLE) perform shallow routing on this already-compressed representation; they can re-weight a fixed set of basis vectors but cannot reconstruct task-aware feature organizations that were never formed during backbone processing. The bottleneck is structural, not a matter of routing capacity.
\item \textbf{Diminishing multi-task returns from scaling}: Scaling depth, width, or sequence length primarily enlarges shared representational capacity; the additional parameters serve task-agnostic feature crossing rather than task-specific signal separation. We empirically observe that a task-agnostic unified backbone converts parameter growth to multi-task GAUC gains at a substantially lower rate than a task-aware alternative, confirming a structural mismatch between generic scaling and multi-task needs.
\end{itemize}

\begin{figure}[t]     
  \centering
  \includegraphics[width=1.0\linewidth]{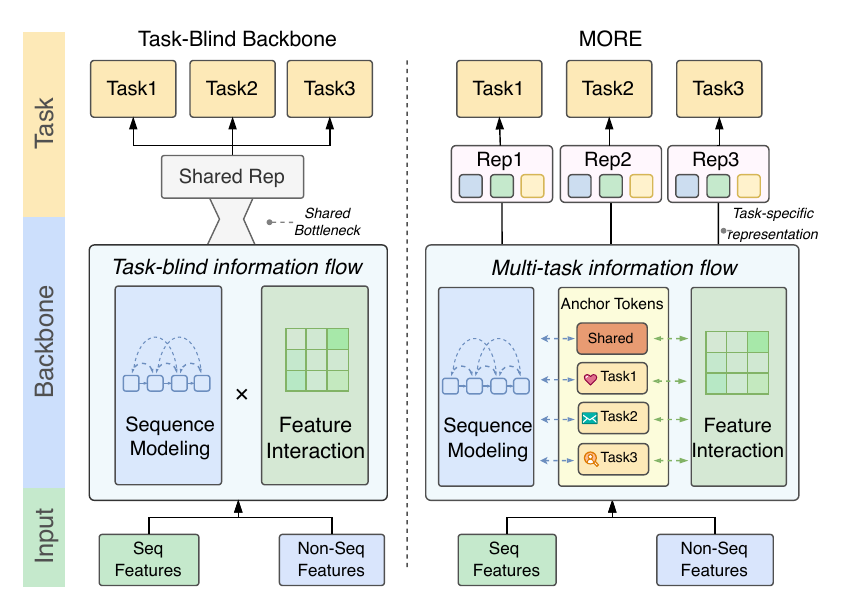}
  \caption{Comparison of information flow:  Task-blind backbone vs MORE.}
  \Description{Architecture diagram of MORE with Anchor Generation, stacked MORE Blocks, and task-specific prediction heads.}
  \label{fig:intro}
\end{figure}

MORE (Multi-task cO-evolving Ranking modEl) is a task-aware backbone for large-scale recommendation ranking. As contrasted in Figure~\ref{fig:intro}, its core design moves multi-task modeling from the prediction stage into the backbone. It introduces Anchor Tokens as task-aware information carriers that pass through all blocks. Shared Anchors encode cross-task semantics, while Private Anchors encode task-specific priors. In each block, Anchor Tokens interact with behavior sequences and non-sequential features in a task-aware manner, and explicit task boundaries reduce cross-task semantic interference. Through stacked blocks, task information and sequence states co-evolve layer by layer, making multi-task modeling a continuous information flow inside the backbone. To meet industrial deployment requirements, MORE also reduces online scoring latency by about 30\% through parallel inference and request aggregation.
Our contributions are summarized as follows:
\begin{itemize}
\item We identify a structural gap in unified ranking backbones: multi-task modeling is confined to post-backbone prediction heads while the backbone itself remains task-agnostic. We analyze the resulting limitations in task-specific information flow, representation capacity, and scaling efficiency.
\item We propose MORE, a task-aware unified backbone that introduces Anchor Tokens, including Shared Anchors for cross-task semantics and Private Anchors for per-task priors, to co-evolve multi-task signals with sequence and feature representations across all backbone layers, with explicit task-boundary isolation in every block. Because the architecture separates user-level sequence states from per-candidate anchor states, it further admits request-level parallel scoring that reduces online latency by about 30\% with no model modification.
\item Offline experiments on large-scale industrial data show that MORE achieves positive GAUC gains across all seven tasks under comparable FLOPs budgets and converts scaling of both parameters and compute into multi-task gains more efficiently than task-agnostic alternatives. Online A/B tests on a social platform with tens of millions of monthly active users report consistent improvements spanning browsing, interaction, and deep social engagement. MORE is deployed in production.
\end{itemize}

\begin{figure*}[t]
  \centering
  \includegraphics[width=0.95\textwidth]{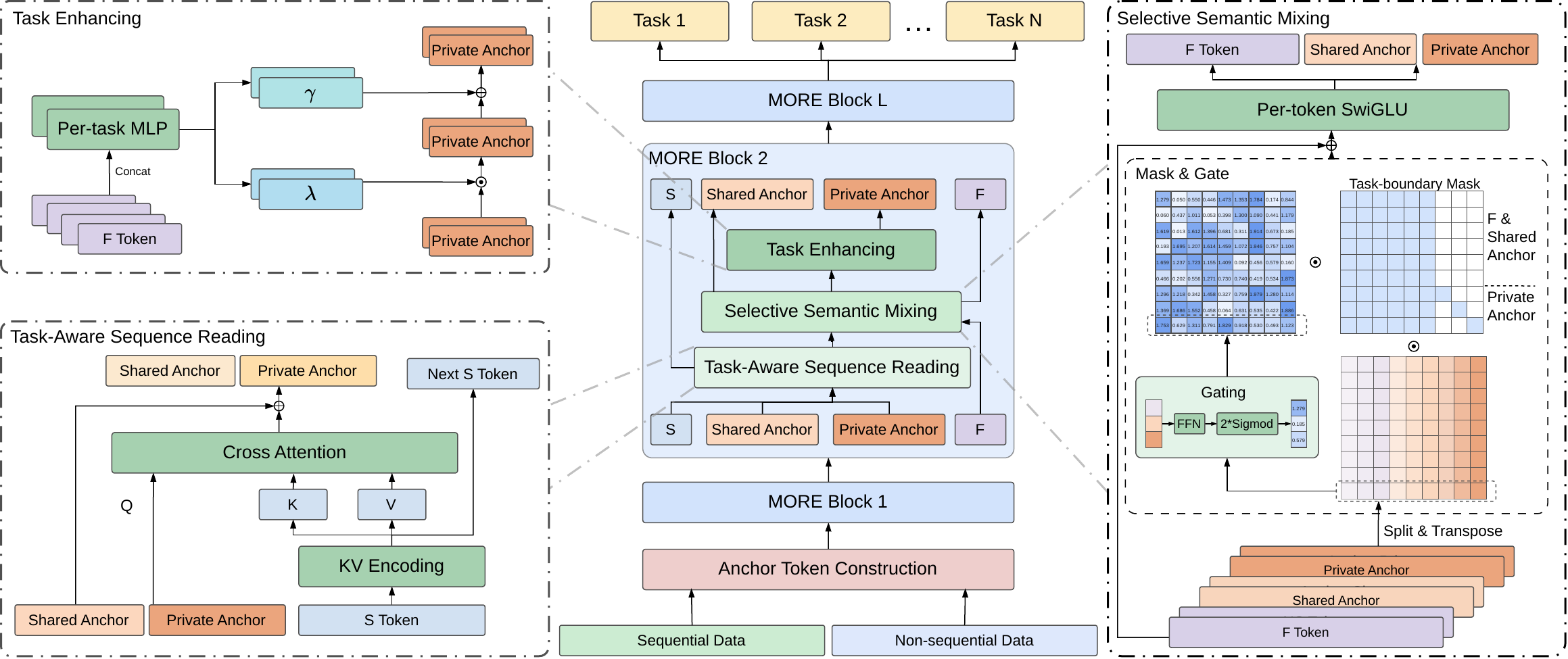}
  \caption{Overview of MORE. MORE consists of Anchor Token Construction, stacked MORE Blocks, and task-specific prediction heads. Each MORE Block contains three modules: Task-Aware Sequence Reading, Selective Semantic Mixing, and Task
Enhancing. Specifically, Task-Aware Sequence Reading strengthens sequential signal modeling, Selective Semantic Mixing
enhances shared semantics and task-boundary awareness, and Task Enhancing improves the task-discriminative capacity of
Private Anchors. This design enables MORE to jointly model shared and task-specific information within a unified backbone.}
  \Description{Architecture diagram of MORE with Anchor Generation, stacked MORE Blocks, and task-specific prediction heads.}
  \label{figoverview}
\end{figure*}

\section{Related Work}
\label{secrelated}

Industrial ranking models follow the DLRM paradigm~\cite{naumov2019dlrm}: classic methods including Wide \& Deep~\cite{cheng2016wide}, DeepFM~\cite{guo2017deepfm}, xDeepFM~\cite{lian2018xdeepfm}, and DCN/DCN-V2~\cite{wang2017dcn,wang2021dcnv2} embed sparse features, model feature interactions through cross networks or MLPs, and attach prediction heads for multi-task scoring. In this feature-centric pipeline, later work extends backbone capacity along two independent routes. Feature interaction methods, including AutoInt~\cite{song2019autoint}, FiBiNET~\cite{huang2019fibinet}, HiFormer~\cite{gui2023hiformer}, InterFormer~\cite{zeng2024interformer}, and more recently Wukong~\cite{zhang2024wukong} and RankMixer~\cite{rankmixer2025}, scale high-order crossing with industrial evidence for deployability and scaling-law behavior~\cite{kaplan2020scaling}. Sequence modeling methods progress from target-attention with DIN~\cite{zhou2018din} and DIEN~\cite{zhou2019dien} through self-attentive architectures~\cite{vaswani2017attention,kang2018sasrec,chen2019bst} and retrieval-based lifetime-scale approaches such as SIM~\cite{pi2020sim} and TWIN~\cite{twin2023} to end-to-end ten-thousand-level designs such as LONGER~\cite{longer2025} and STCA~\cite{guan2025stca}, and generative transduction with HSTU~\cite{zhai2024hstu}, substantially improving long-term interest extraction. Despite this progress, the two routes remain decoupled in production pipelines, yielding late fusion of heterogeneous signals, and multi-task modeling remains confined to post-backbone prediction heads.

To bridge this separation, recent unified architectures place sequence modeling and feature interaction within a single stackable backbone. HyFormer~\cite{hyformer2026} introduces Global Tokens as a semantic interface between sequential and non-sequential signals, alternating cross-attention over long sequences with MLP-Mixer~\cite{tolstikhin2021mlpmixer}-style token mixing for bidirectional information flow. MixFormer~\cite{mixformer2026} identifies the co-scaling challenge, where dense capacity and sequence capacity compete for the same compute budget, and unifies both within a single parameterization that exhibits better scaling behavior over FLOPs and sequence length. OneTrans~\cite{onetrans2025} maps both signal types into a common token space through a unified tokenizer and progressively compresses sequence information via pyramid stacking. These designs improve accuracy and scalability, yet their backbones remain task-agnostic: task preferences over sequences and features are expressed only by prediction heads, while sequence reading and feature interaction inside the backbone lack awareness of per-task distinctions.

Multi-task learning in ranking has advanced through foundational frameworks such as MMoE~\cite{ma2018mmoe}, which uses mixture-of-experts with task-specific gating, and PLE~\cite{tang2020ple}, which applies progressive expert separation against negative transfer, and subsequent refinements such as AITM~\cite{xi2021aitm} and PEPNet~\cite{chang2023pepnet}, yet all confine task differentiation to post-backbone towers, leaving the backbone itself to produce a task-agnostic shared representation. Recent attempts to inject task signals into earlier backbone layers~\cite{infnet2025} still share interaction parameters across all tasks and leave behavior-sequence updates undifferentiated per task, so that per-task sequence reading and explicit cross-task isolation within the backbone remain unaddressed in unified ranking architectures.

\section{Method}

\subsection{Problem Statement}

Let $\mathcal{U}$ and $\mathcal{C}$ denote the user space and item space respectively. For a given user $u \in \mathcal{U}$, the behavior history consists of a sequence with multiple behavior types, and the candidate set is $\{c_1, \ldots, c_C\}$. Let $\mathbf{S}$ denote the user's historical behavior sequence features. Non-sequential features $\mathbf{NS}$ comprise user profiles, candidate attributes, and context features. The system is required to output prediction scores for $T$ tasks, such as click, like, comment, greet, and chat. Formally, the ranking model outputs:
\begin{equation}
\hat{y}_1,\ldots,\hat{y}_T = f(\mathbf{S},\;\mathbf{NS};\,\Theta).
\end{equation}

The model parameters $\Theta$ are optimized by minimizing the weighted binary cross-entropy loss, where $\lambda_t$ is the task-specific loss weight:
\begin{equation}
\mathcal{L} = \sum_{t=1}^{T} \lambda_t\mathcal{L}_{\mathrm{BCE}}(\hat{y}_t, y_t).
\end{equation}

\subsection{Framework Overview}

As shown in Figure~\ref{figoverview}, MORE comprises an input representation layer, $L$ stackable MORE blocks, and $T$ prediction towers. Each MORE block contains three submodules that update Anchor Tokens sequentially: Task-Aware Sequence Reading extracts task-aware signals from behavior sequences via cross-attention; Selective Semantic Mixing enables global semantic interaction between Anchor Tokens and $\mathbf{F}$, with a task-boundary mask applied to isolate task boundaries; Task Enhancing embeds discriminative information into each Private Anchor through task-specific independent branches. After stacking $L$ such blocks, each task tower receives the corresponding Private Anchor together with shared semantics from Shared Anchors and non-sequential tokens for multi-task prediction. MORE uses request-level shared computation to maintain industrial deployment efficiency.

\subsection{Input Representation}

MORE organizes ranking-request signals into two token types: sequential tokens encoding multi-behavior history, and non-sequential tokens encoding user profiles, candidate attributes, and context for each candidate, both mapped into a shared $d$-dimensional space. Anchor Tokens are further derived from these inputs to carry multi-task information as explicit states through the backbone.

\paragraph{Task-Aware Multi-Behavior Sequence Encoding.}
User behavior history comprises $T$ types (click, interact, chat, etc.). Encoding only item-level positions without behavior-type discrimination weakens per-task structure in later layers. MORE applies Cartesian product encoding: each item's behavior combination is represented as a length-$T$ binary indicator mapped to a unique action combination ID, and each item--ID pair yields a distinct sequence token
\begin{equation}
\mathbf{s}_n = \mathbf{e}_{\mathrm{item}}(i_n) + \mathbf{e}_{\mathrm{action}}(\mathrm{cid}_n) + \mathbf{e}_{\mathrm{time}}, \quad n= 1, \ldots, N,
\end{equation}
where $\mathbf{e}_{\mathrm{item}}(i_n)$ is the embedding of item $i_n$, $\mathbf{e}_{\mathrm{action}}(\mathrm{cid}_n)$ is the embedding of the action combination ID, and $\mathbf{e}_{\mathrm{time}}$ is a learnable positional embedding encoding temporal order. $\mathbf{S} \in \mathbb{R}^{N \times d}$ separates behavior-type combinations into distinct tokens, providing fine-grained input for Task-Aware Sequence Reading and mitigating cross-task gradient conflicts at each position.

\paragraph{Non-Sequential Semantic Tokens.}
For each candidate, user profiles, candidate attributes, and context features are embedded and grouped into a non-sequential token matrix $\mathbf{F} \in \mathbb{R}^{M \times d}$. In the backbone, $\mathbf{F}$ provides contextual semantics for Anchor Tokens during Selective Semantic Mixing and injects discriminative information into Private Anchors during Task Enhancing.

\paragraph{Anchor Token Construction.}
MORE introduces Anchor Tokens as cross-module carriers of multi-task information through all backbone layers. They are initialized by concatenating a mean-pooled behavior summary from $\mathbf{S}$ with flattened non-sequential semantics from $\mathbf{F}$, then projecting and splitting via an MLP:

\begin{equation}
\begin{gathered}
\mathbf{h} = \mathrm{Concat}\!\big(\mathrm{MeanPool}(\mathbf{S}),\; \mathrm{Flatten}(\mathbf{F})\big), \\
\mathbf{A}_1,\ldots, \mathbf{A}_{K+T}= \mathrm{Split}\big(\mathrm{MLP}(\mathbf{h})\big).
\end{gathered}
\end{equation}

The first $K$ outputs form Shared Anchors $\mathbf{A}_s \in \mathbb{R}^{K \times d}$ for cross-task semantics; the remaining $T$ form Private Anchors $\mathbf{A}_p \in \mathbb{R}^{T \times d}$, one per task. A learnable task prior embedding $\boldsymbol{\pi}_t \in \mathbb{R}^d$ is added to strengthen task discrimination:
\begin{equation}
\mathbf{A}_p^{(t)} = \mathbf{A}_p^{(t)} + \boldsymbol{\pi}_t, \quad t = 1, \ldots, T.
\end{equation}

\subsection{MORE Block}

\subsubsection{Task-Aware Sequence Reading}

The same behavior history often carries task-dependent discriminative signals. Task-Aware Sequence Reading lets Anchor Tokens read the shared behavior sequence at different levels, with Shared Anchors reading task-shared general information and Private Anchors reading task-relevant signals.

Given the behavior sequence $\mathbf{S} \in \mathbb{R}^{N \times d}$ and Anchor Tokens $\mathbf{A}=[\mathbf{A}_s;\,\mathbf{A}_p]$, each block updates the shared sequence and forms the current layer's cross-attention key-value states:
\begin{equation}
\tilde{\mathbf{S}}=\mathrm{SwiGLU}(\mathbf{S}),\quad \mathbf{K}=\tilde{\mathbf{S}}\mathbf{W}_{K},\quad \mathbf{V}=\tilde{\mathbf{S}}\mathbf{W}_{V}.
\end{equation}
Here $\mathbf{W}_{K},\mathbf{W}_{V}\in\mathbb{R}^{d\times d}$ are key and value projection matrices, and $\mathbf{K},\mathbf{V}\in\mathbb{R}^{N\times d}$ denote the resulting sequence memory.

The updated sequence $\tilde{\mathbf{S}}$ is forwarded to the next layer, so sequence representations evolve with block depth. To provide Anchor Tokens with cross-block sequence memory, MORE maintains a sequence register $\boldsymbol{\eta}\in\mathbb{R}^{d}$, initialized to zero. At the end of each block, $\boldsymbol{\eta}$ accumulates a compressed summary of the current-layer sequence:
\begin{equation}
\boldsymbol{\eta} \leftarrow \mathrm{LN}\!\big(\boldsymbol{\eta}+\mathrm{MLP}(\mathrm{Pool}(\tilde{\mathbf{S}}))\big).
\end{equation}

Anchor Tokens then query the shared sequence memory. After cross-attention, each Shared Anchor and each Private Anchor is augmented by the register through its own independent SwiGLU gate, so that every position interprets the shared sequence summary differently:
\begin{equation}
\hat{\mathbf{A}}_s^{(k)} =\mathrm{CrossAttn}(\mathbf{A}_s^{(k)},\mathbf{K},\mathbf{V}) + \mathrm{SwiGLU}_k(\boldsymbol{\eta}), \quad k = 1, \ldots, K.
\end{equation}
\begin{equation}
\hat{\mathbf{A}}_p^{(t)} =\mathrm{CrossAttn}_{p}^{(t)}(\mathbf{A}_p^{(t)},\mathbf{K},\mathbf{V}) + \mathrm{SwiGLU}_t(\boldsymbol{\eta}), \quad t = 1, \ldots, T.
\end{equation}
Here $\mathrm{SwiGLU}_k$ and $\mathrm{SwiGLU}_t$ are per-anchor and per-task gated projections respectively. The shared register captures global sequence evolution across blocks, while independent gating allows each anchor to selectively extract the signals most relevant to its role.

\subsubsection{Selective Semantic Mixing}

After sequence reading, Anchor Tokens have incorporated sequence information, but their interaction with non-sequential tokens $\mathbf{F}$ has not yet been established. If all tokens are fused without differentiation during mixing, task-specific semantics of different Private Anchors will dilute one another. Inspired by RankMixer~\cite{rankmixer2025}, Selective Semantic Mixing applies a lightweight MLP-Mixer-style token-mixing operation to enrich sequence-aware anchors with non-sequential semantics while preserving task boundaries.

The sequence-reading output and non-sequential tokens $\mathbf{F}$ are concatenated into a unified token matrix
\begin{equation}
\mathbf{G} = (\mathbf{F};\, \hat{\mathbf{A}}_s;\, \hat{\mathbf{A}}_p^{(1)},\, \ldots,\, \hat{\mathbf{A}}_p^{(T)}) \in \mathbb{R}^{H \times d},
\end{equation}

where $H = M + K + T$ and $d_h = d/H$. The embedding dimension of $\mathbf{G}$ is split into $H$ heads of size $d_h$, and the head and token axes are transposed, yielding a three-dimensional tensor $\tilde{\mathbf{G}} \in \mathbb{R}^{H \times H \times d_h}$ whose axes are (head, token, subspace). The $i$-th head slice $\tilde{\mathbf{g}}_i = \tilde{\mathbf{G}}_{i,:,:} \in \mathbb{R}^{H \times d_h}$ collects the representations of all $H$ tokens in the $i$-th subspace. To adaptively calibrate the importance of each channel, each slice passes through an MLP gate to produce per-token scores:
\begin{equation}
m_i = \mathrm{MLP}_i(\tilde{\mathbf{g}}_i) \in \mathbb{R}^{H}, \quad \mathrm{Gate}= (m_1,\, \ldots,\, m_H) \in \mathbb{R}^{H \times H}.
\end{equation}

To ensure that all Private Anchors are strictly invisible to each other throughout the entire iteration process, MORE constructs a task-boundary mask $\mathbf{M}\in \{0, 1\}^{H \times H}$. As shown in Figure~\ref{figoverview}: $\mathbf{F}$ and Shared Anchors are visible to all tokens, while each Private Anchor is only visible to itself. For each head $i$, the gate vector $m_i$ and the $i$-th row of the mask $\mathbf{M}_{i,:}$ are multiplied element-wise to form a joint weight $\in \mathbb{R}^{H}$, which is broadcast along $d_h$ and applied to the head slice; all gated slices are then transposed back and concatenated across heads to recover the token layout:
\begin{equation}
\begin{gathered}
\hat{\mathbf{g}}_i = \tilde{\mathbf{g}}_i \odot (m_i \odot \mathbf{M}_{i,:}) \in \mathbb{R}^{H \times d_h}, \quad i = 1, \ldots, H, \\
\hat{\mathbf{G}} = \mathrm{Reshape}\!\big(\hat{\mathbf{g}}_1, \ldots, \hat{\mathbf{g}}_H\big) \in \mathbb{R}^{H \times d},
\end{gathered}
\end{equation}
where $\mathrm{Reshape}$ concatenates all head slices along the embedding dimension to recover $\mathbb{R}^{H\times d}$.

Global semantic fusion is then performed through two per-token FFN layers and residual connections:
\begin{equation}
\begin{gathered}
\mathbf{G}_{\mathrm{mid}} = \mathrm{Norm}\!\big(\mathrm{PFFN}_1(\hat{\mathbf{G}}) + \mathbf{G}\big), \\
\mathbf{G}_{\mathrm{out}} = \mathrm{Norm}\!\big(\mathrm{PFFN}_2(\mathbf{G}_{\mathrm{mid}}) + \mathbf{G}_{\mathrm{mid}}\big).
\end{gathered}
\end{equation}

Together with the task-boundary mask, PFFN allows global semantic flow without crossing task boundaries. The output $\mathbf{G}_{\mathrm{out}}$ is then split along the token axis, reversing Eq.~(10):
\begin{equation}
(\mathbf{F},\; \mathbf{A}_s,\; \bar{\mathbf{A}}_p^{(1)}, \ldots, \bar{\mathbf{A}}_p^{(T)}) = \mathrm{Split}(\mathbf{G}_{\mathrm{out}}).
\end{equation}
\subsubsection{Task Enhancing}

The mixing stage applies the same global mixing to all tokens. As task-specific anchors for downstream prediction, Private Anchors require richer task-specific semantics. Inspired by FiLM~\cite{perez2018film}, Task Enhancing injects high-dimensional discriminative information from $\mathbf{F}$ into each Private Anchor through independent per-task enhancement branches, expanding the semantic capacity of task-specific representations. For the $t$-th task, an independent MLP aggregates conditioning parameters from non-sequential features:
\begin{equation}
(\boldsymbol{\lambda}_t,\; \boldsymbol{\gamma}_t) = \mathrm{Split}\!\big(\mathrm{MLP}_t(\mathrm{Concat}(\mathbf{F}))\big), \quad \boldsymbol{\lambda}_t, \boldsymbol{\gamma}_t \in \mathbb{R}^d.
\end{equation}

The Private Anchor is then updated via FiLM-style affine modulation with a residual connection:
\begin{equation}
\mathbf{A}_p^{(t)}=\mathrm{Norm}\!\big((\bar{\mathbf{A}}_p^{(t)} \odot \boldsymbol{\lambda}_t + \boldsymbol{\gamma}_t) + \bar{\mathbf{A}}_p^{(t)}\big).
\end{equation}

\subsubsection{Block Stacking}
$L$ MORE blocks are stacked sequentially, and the Reading, Mixing and Enhancing submodules within each block recurrently update hidden states. The complete recurrence formulation at layer $l$ is:
\begin{equation}
\begin{gathered}
(\hat{\mathbf{A}}^{(l)}, \mathbf{S}^{(l)}) = \mathrm{Reading}\!\big(\mathbf{A}^{(l-1)},\; \mathbf{S}^{(l-1)}\big), \\
(\mathbf{F}^{(l)},\; \mathbf{A}_s^{(l)},\; \bar{\mathbf{A}}_p^{(l)}) = \mathrm{Mixing}\!\big(\mathbf{F}^{(l-1)},\; \hat{\mathbf{A}}^{(l)}\big), \\
\mathbf{A}_p^{(l)} = \mathrm{Enhancing}\!\big(\mathbf{F}^{(l)},\; \bar{\mathbf{A}}_p^{(l)}\big).
\end{gathered}
\end{equation}

After passing through $L$ MORE blocks, $\mathbf{F}$ and $\mathbf{A}_s$ act as cross-task global representations, while $\mathbf{A}_p^{(t)}$ encodes task-specific semantic features. Each task tower receives the corresponding Private Anchor together with shared semantics from $\mathbf{F}$ and $\mathbf{A}_s$, retaining cross-task context while preserving task-specific prediction.

\subsection{Request-Level Training and Inference}
\label{secrlb}

In industrial ranking, a single request typically involves thousands of candidates. Under per-candidate deployment, each candidate is forwarded independently, causing user-side sequence encoding and shared feature computation to repeat $C$ times.

Inspired by the request-level organization in HSTU~\cite{zhai2024hstu}, MORE adopts request-level shared computation to unify sample organization across training and inference. The $C$ candidates under one request are aggregated into a single sample: the user-side sequence $\mathbf{S}$ is encoded once and broadcast, while per-candidate $\mathbf{F}$ and Anchor Tokens are organized along the item dimension. At inference time, candidates are split into mini-batches, and within each mini-batch user-side computation is shared across all candidates.

This organization aligns with the internal state structure of MORE Blocks. $\mathbf{S}$ is a user-level shared state that evolves through layer-wise transformations without depending on candidate information (Eq.~6), while per-candidate differences are expressed through $\mathbf{F}$ and Anchor Tokens and do not participate in the sequence forward computation. Request-level shared computation therefore requires no architectural adaptation of the backbone, reducing online scoring latency by about 30\%.

\section{Experiment}

\begin{table*}[t]
  \caption{Offline GAUC comparison on seven tasks. Transformer+RankMixer reports absolute GAUC, and other rows report relative change. Params denote dense parameters; FLOPs denote forward-pass compute per example.}
  \label{taboverall}
  \centering
  \small
  \setlength{\tabcolsep}{4pt}
  \begin{tabular}{@{}l c c c c c c c c c c@{}}
    \toprule
    \textbf{Method} & \textbf{Category} & \textbf{Click} & \textbf{Like} & \textbf{Comment} & \textbf{Greet} & \textbf{Reply} & \textbf{Avatar-click} & \textbf{Deep Chat} & \textbf{Params} & \textbf{FLOPs/Ex.} \\
    \midrule
    Transformer+RankMixer & Two-Stage & 0.7863 & 0.6892 & 0.6979 & 0.7279 & 0.8388 & 0.7073 & 0.8444 & 77.85M & 1.92G \\
    Transformer+Wukong & Two-Stage & +0.03\% & -0.27\% & -0.65\% & -0.53\% & -0.15\% & -0.36\% & -0.46\% & 79.01M & 1.90G \\
    HSTU+RankMixer & Two-Stage & +0.09\% & -0.23\% & -0.57\% & -0.34\% & -0.10\% & -0.25\% & -0.24\% & 77.19M & 1.76G \\
    Transformer+DCN-v2 & Two-Stage & +0.12\% & +0.53\% & -0.04\% & -0.02\% & -0.16\% & +0.17\% & -0.47\% & 111.78M & 1.68G \\
    STCA & Two-Stage & +0.13\% & +0.66\% & +0.38\% & -0.01\% & +0.02\% & +0.06\% & -0.07\% & 88.21M & 2.36G \\
    MixFormer & Unified & +0.29\% & +1.13\% & +0.89\% & +1.25\% & +0.49\% & +0.70\% & +0.15\% & 98.68M & 4.65G \\
    HyFormer & Unified & +0.43\% & +0.89\% & +1.09\% & +1.32\% & +0.42\% & +0.82\% & +0.08\% & 153M & 2.16G \\
    OneTrans & Unified & +0.38\% & +1.15\% & +0.80\% & +1.44\% & +0.58\% & +0.86\% & +0.11\% & 108.65M & 1.73G \\
    MORE & Task-Aware Unified & \textbf{+0.44\%} & \textbf{+1.47\%} & \textbf{+1.26\%} & \textbf{+1.58\%} & \textbf{+0.62\%} & \textbf{+1.02\%} & \textbf{+0.33\%} & 132M & 1.85G \\
    \bottomrule
  \end{tabular}
\end{table*}

\subsection{Experimental Setup}

\subsubsection{Dataset}
The offline experiments are conducted in Nearby Feed, a social multimedia feed on Momo where users discover posts from nearby strangers and connect with those who interest them. The dataset is derived from online recommendation logs spanning 90 consecutive days for training and the immediately following day for evaluation.
Input features include sequential and non-sequential features. Sequential features record the user's recent multi-type behavior history. Non-sequential features include user profiles, candidate content attributes, and scenario context.

\subsubsection{Tasks and Metrics}
The model jointly optimizes thirteen tasks covering browsing, interaction, and social engagement, including click, like, comment, follow, greet, reply, avatar-click, and deep chat (conversations reaching at least five rounds), among others. We use GAUC (Group AUC) as the primary offline metric, which computes AUC per user and averages across users to reflect individual-level ranking quality.

\subsubsection{Baselines}
Our baselines span two paradigms. Two-stage baselines separate sequence encoding from feature interaction: the reference uses Transformer + RankMixer~\cite{rankmixer2025}; variants swap the feature-interaction module for Wukong~\cite{zhang2024wukong} or DCN-v2~\cite{wang2021dcnv2}, and the sequence encoder for HSTU~\cite{zhai2024hstu}. STCA~\cite{guan2025stca} is listed separately as its architecture integrates sequence encoding and RankMixer-style feature interaction into one pipeline. Unified-backbone baselines include MixFormer~\cite{mixformer2026}, HyFormer~\cite{hyformer2026}, and OneTrans~\cite{onetrans2025}; their architectures are described in Section~\ref{secrelated}. All methods are reimplemented in the same framework with identical training data, schedule, evaluation protocol, and hyperparameter budget.

\subsubsection{Implementation Details}
MORE uses $L_{\text{block}} = 2$ MORE Blocks by default, hidden dimension $d = 256$, $M = 15$ non-sequential tokens in $\mathbf{F}$, and 17 Anchor Tokens, including 8 Shared Anchors and 9 Private Anchors corresponding to the core tasks. The model is optimized with Adam using a batch size of 1024 and trained on a multi-GPU data-parallel cluster. Section~\ref{secscaling} scales layers and hidden dimension, expanding parameters from about 0.1B to about 0.9B and forward FLOPs from about 1.4G to about 11G per example, to study MORE's scaling behavior.

\subsection{Overall Performance}

The main results for MORE and the baselines are summarized in Table~\ref{taboverall}. We take a closer look at each model group. Separately swapping the feature-interaction or sequence-modeling component yields limited and unstable gains for multi-task ranking: replacing RankMixer with Wukong or DCN-v2 degrades most tasks---DCN-v2 hurts Deep Chat by -0.47\%---confirming RankMixer's advantage as a feature-interaction backbone, while replacing Transformer with HSTU for sequence encoding benefits only Click and degrades all other tasks. STCA integrates sequence encoding and feature interaction into a single pipeline, improving interaction-level tasks (Like +0.66\%, Comment +0.38\%) over the fully separated baselines, but its backbone remains task-agnostic and deeper social tasks such as Greet and Deep Chat show negligible or negative change. Unified sequence-feature backbones further alleviate this limitation by placing sequential and non-sequential features in one backbone: MixFormer, HyFormer, and OneTrans all improve on every task, with OneTrans reaching +1.44\% on Greet and HyFormer +1.09\% on Comment. However, the improvement pattern across tasks is similar for the three methods, consistent with their task-agnostic backbones that differentiate tasks only at prediction heads. As a result, these methods reduce sequence-feature separation but do not provide differentiated task-level information flow during sequence reading and semantic interaction.

MORE further moves multi-task information flow into the backbone. It leads all tasks in Table~\ref{taboverall} (e.g., Greet +1.58\%, Like +1.47\%, Deep Chat +0.33\%), with particularly strong incremental gains on interaction and social tasks where behavior-sequence signals require task-specific reading. Shared Anchors read cross-task evidence, Private Anchors use task-conditioned queries for task-specific signals, and semantic mixing preserves task boundaries through the task-boundary mask. All tasks benefit, confirming that task-aware anchors and task boundaries improve multi-task ranking beyond what a task-agnostic shared representation provides.

MORE also achieves a favorable accuracy-efficiency trade-off. Despite introducing task-aware capacity that increases parameters to 132M (still below HyFormer's 153M), its per-example FLOPs (1.85G) are lower than the reference baseline (1.92G) and substantially lower than STCA (2.36G), HyFormer (2.16G), and MixFormer (4.65G). Compared with OneTrans, which has lower FLOPs (1.73G), MORE adds only 0.12G per example through per-task query projections and task-boundary isolation while outperforming it on all tasks. This efficiency comes from injecting multi-task capacity through a compact set of Anchor Tokens rather than duplicating the backbone or substantially increasing token mixing.

\subsection{Ablation Study}

\begin{table}[t]
  \centering
  \caption{Ablation study on MORE components. Interaction denotes the average GAUC of like, comment, and greet. Average is the arithmetic mean of the Click, Interaction, and Chat columns.}
  \label{tabablation}
  \small  
  \setlength{\tabcolsep}{3pt} 
  \begin{tabular}{lcccc}
    \toprule
    \textbf{Configuration} 
    & \textbf{Click}
    & \textbf{Interaction}
    & \textbf{Chat} 
    & \textbf{Average}\\
    \midrule
    \multicolumn{5}{l}{\textbf{Ablation of Token Process}}\\
    \midrule

    w/o Cartesian Encoding      & -0.06\% & -0.10\% & -0.08\% &-0.08\%\\
    w/o Anchor From Context       & -0.08\% & -0.04\% & -0.17\% &-0.10\%\\
    \midrule
    \multicolumn{5}{l}{\textbf{Ablation of Seq Reading}}\\
    \midrule
    w/o Private Seq Reading       & -0.15\% & -0.19\% & -0.12\% & -0.15\%\\  
    w/o Seq Register  & -0.02\% & -0.01\% & -0.13\% & -0.05\%\\
    \midrule
    \multicolumn{5}{l}{\textbf{Ablation of Mixing and Enhancing}}\\
    \midrule
    w/o Boundary Gating Mask  & -0.16\% & -0.34\% & -0.11\% & -0.21\%\\
    w/o Task Enhancing  & -0.11\% & -0.13\% & -0.17\% &-0.14\% \\
    \midrule
    \multicolumn{5}{l}{\textbf{Ablation of all Components}}\\
    \midrule
    BaseArch  & -0.23\% & -0.38\% & -0.39\% & -0.33\%\\
    \bottomrule
  \end{tabular}
\end{table}

Table~\ref{tabablation} shows ablation results for core components of MORE. Components are removed by module group and relative changes in GAUC on click, interaction (average of like, comment, and greet), deep chat, and their average are reported.

The two Token Process components enable the model to acquire multi-task awareness from the initial stage, and their removal leads to consistent GAUC drops of 0.04\%--0.10\% across reported tasks.
Private Seq Reading leverages per-task independent parameters to boost behavioral sequence perception for each private anchor. Seq Register establishes sequence registers, maintaining low-level sequence details. Both components enhance the multi-task capability during the reading stage. Boundary Gating Mask isolates information flow among different Private Anchors in Eq.~(12), preventing semantic contamination across tasks while preserving global semantic interaction. Task Enhancing injects discriminative information into each Private Anchor through lightweight independent per-task branches, contributing 0.11\%--0.17\% GAUC improvement across tasks.
Among individual components, Boundary Gating Mask causes the largest average degradation when removed, with a 0.34\% drop on Interaction. Without the mask, gradient signals from high-frequency tasks dominate shared mixing weights, diluting the representations that interaction and social tasks rely on. Task-boundary control in the mixing stage thus proves more critical than per-task capacity in either the reading or enhancing stage alone.

When all six components are removed, the model degenerates to BaseArch, a task-agnostic unified backbone that retains only multiple prediction heads for multi-task scoring, and average GAUC decreases by 0.33\%. This result shows the cumulative contribution of the components to final performance.

\subsection{Scaling Analysis}
\label{secscaling}

\begin{figure}[t]
  \centering
  \includegraphics[width=\linewidth]{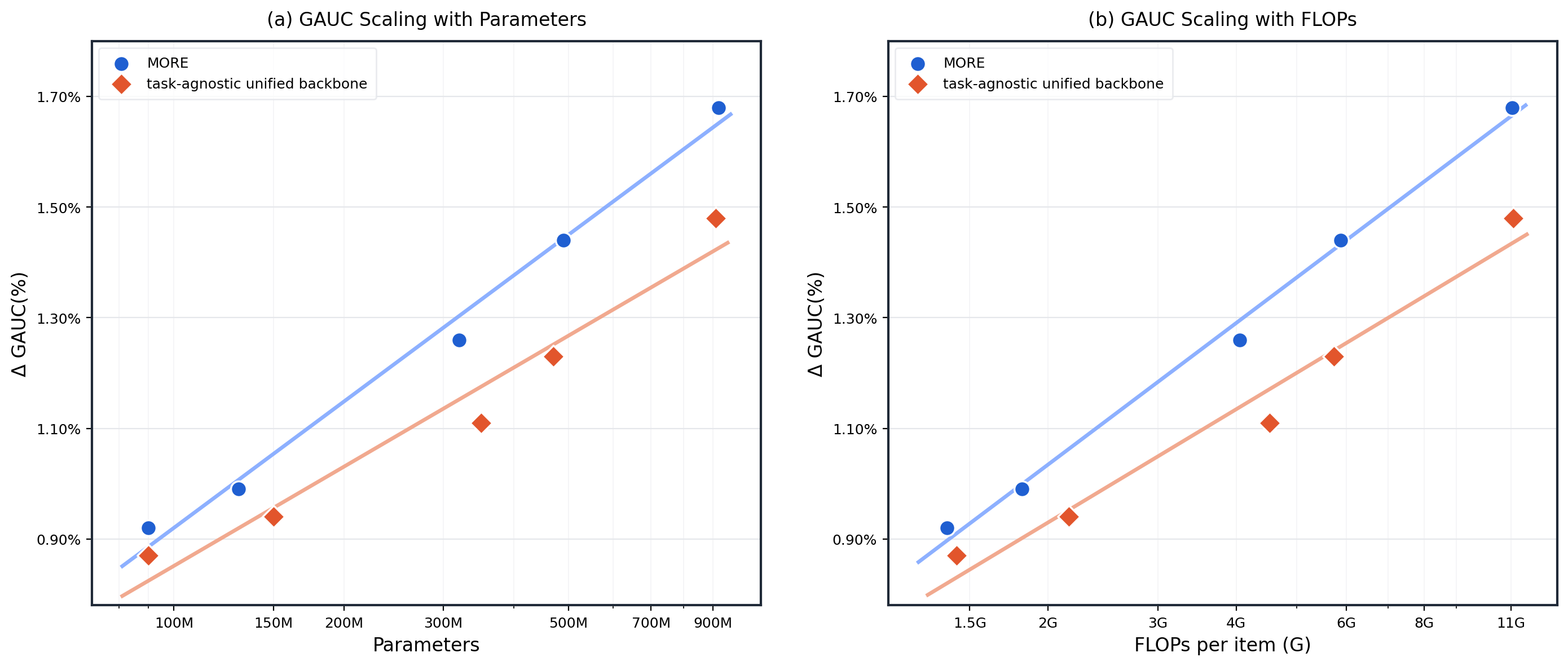}
  \caption{Scaling comparison on average GAUC improvement relative to the online production model. (a) GAUC vs.\ parameters; (b) GAUC vs.\ per-example forward FLOPs.}
  \Description{Two-panel scatter plot comparing MORE and a task-agnostic unified backbone in average GAUC improvement relative to the online production model, plotted against parameters (left) and FLOPs (right), with fitted trend lines.}
  \label{fig:scaling_params_flops}
\end{figure}

To examine whether the task-aware mechanism of MORE remains effective under different model capacities, the scaling study jointly adjusts the number of MORE Blocks and the hidden dimension, covering model sizes from about 0.1B to about 0.9B parameters. The task-agnostic baseline is a unified backbone that alternates sequence cross-attention with token mixing following the HyFormer~\cite{hyformer2026} design, scaled to the same parameter ranges but without task-aware components. Each configuration is trained independently on the same data and evaluated under the same protocol.
Figure~\ref{fig:scaling_params_flops} reports the average GAUC improvement across all tasks relative to the online production model, plotted against both parameter count and per-example forward FLOPs.

The fitted trends show that MORE scales more effectively across both parameters and compute, remaining above the task-agnostic unified backbone at comparable budgets. The task-agnostic baseline also improves with scale, but its slower growth suggests that additional capacity and compute are converted less efficiently into multi-task GAUC gains. The consistent advantage across all tested scales confirms that the task-aware capacity introduced by Anchor Tokens is productively leveraged at every model size, converting additional parameters and compute into multi-task gains more efficiently than the task-agnostic design. At the smallest configuration both approaches yield comparable GAUC; as capacity grows the gap widens, confirming that task-agnostic scaling alone cannot substitute for per-task signal separation.

\subsection{Online A/B Test}

MORE is evaluated in the Nearby Feed recommendation scenario of Momo against the production DLRM-style ranking model. A two-week online A/B experiment is conducted with balanced traffic splitting; metrics remain stable throughout the experiment with no degradation trend. Table~\ref{tabonline} reports relative improvements on six primary online evaluation metrics---usage duration, click, like, comment, greet, and deep chat---spanning session-level engagement, shallow interaction, and deep social connection. All reported improvements are statistically significant (p<0.05).

MORE achieves positive improvements across all six metrics. Usage duration increases by 3.0\%, reflecting overall improvement in user engagement within the social discovery scenario. Click (+2.8\%), like (+2.0\%), and comment (+3.6\%) indicate that the model surfaces more relevant content, improving interaction quality. Greet (+2.5\%) and deep chat (+2.0\%) represent stronger social matching---the model surfaces people that users find more appealing and are more likely to sustain conversations with, demonstrating improved multi-task ranking across content interaction and social connection.

With the request-level shared computation strategy described in Section~\ref{secrlb}, MORE reduces P99 scoring latency by approximately 30\% compared to its per-candidate serving baseline, satisfying the latency constraints of high-concurrency recommendation serving. The reduction stems from sharing user-side sequence computation across candidates within each mini-batch, effectively amortizing the per-candidate cost of the unified backbone. MORE is deployed in production and serves online traffic in this scenario.

\begin{table}[t]
  \centering
  \caption{Online A/B test results on Momo Nearby Feed over two weeks. MORE replaces the production ranking model.}
  \label{tabonline}
  \small
  \begin{tabular}{@{}lc@{}}
    \toprule
    \textbf{Metric} & \textbf{Relative Lift} \\
    \midrule
    Usage Duration & +3.0\% \\
    Click          & +2.8\% \\
    Like           & +2.0\% \\
    Comment        & +3.6\% \\
    Greet          & +2.5\% \\
    Deep Chat      & +2.0\% \\
    \bottomrule
  \end{tabular}
\end{table}

\section{Conclusion}

Recent unified backbones merge sequence modeling with feature interaction, yet multi-task modeling remains confined to post-backbone towers. MORE embeds multi-task information flow into the backbone through Shared and Private Anchor Tokens that persist across all blocks. Each block performs task-aware sequence reading, selective semantic mixing under a task-boundary mask, and per-task enhancement, so task signals co-evolve with backbone states at every layer.

Offline, MORE leads all seven tasks against both two-stage and unified baselines. Scaling from 0.1B to 0.9B parameters, MORE converts additional capacity into multi-task GAUC gains more efficiently than task-agnostic alternatives. Online A/B tests yield consistent improvements from browsing to deep social engagement; with request-level shared computation, P99 latency drops about 30\%. MORE is deployed in production on Momo.

The principle of task-aware information flow inside a stackable backbone is not inherently scenario-specific; future work will test generalization across diverse recommendation settings, larger parameter scales, and dynamic task configurations.

\bibliographystyle{ACM-Reference-Format}
\bibliography{references}


\begin{thebibliography}{33}


\ifx \showCODEN    \undefined \def \showCODEN     #1{\unskip}     \fi
\ifx \showISBNx    \undefined \def \showISBNx     #1{\unskip}     \fi
\ifx \showISBNxiii \undefined \def \showISBNxiii  #1{\unskip}     \fi
\ifx \showISSN     \undefined \def \showISSN      #1{\unskip}     \fi
\ifx \showLCCN     \undefined \def \showLCCN      #1{\unskip}     \fi
\ifx \shownote     \undefined \def \shownote      #1{#1}          \fi
\ifx \showarticletitle \undefined \def \showarticletitle #1{#1}   \fi
\ifx \showURL      \undefined \def \showURL       {\relax}        \fi
\providecommand\bibfield[2]{#2}
\providecommand\bibinfo[2]{#2}
\providecommand\natexlab[1]{#1}
\providecommand\showeprint[2][]{arXiv:#2}
\makeatletter
\@ifundefined{NAT@parse@date}{}{\let\NAT@parse@date@orig\NAT@parse@date}
\@ifundefined{NAT@parse@date}{}{\def\NAT@parse@date#1#2#3#4#5#6@@{\NAT@parse@date@orig#1#2#3#4#5#6@@\def\NAT@tempyear{0000}\def\NAT@tempexlab{{?}}\ifx\NAT@year\NAT@tempyear\ifx\NAT@exlab\NAT@tempexlab\def\NAT@date{[n.\,d.]}\else\edef\NAT@date{[n.\,d.]\NAT@exlab}\fi\fi}}
\makeatother

\bibitem[Chai et~al\mbox{.}(2025)]%
        {longer2025}
\bibfield{author}{\bibinfo{person}{Zheng Chai}, \bibinfo{person}{Qin Ren},
  \bibinfo{person}{Xijun Xiao}, \bibinfo{person}{Huizhi Yang},
  \bibinfo{person}{Bo Han}, \bibinfo{person}{Sijun Zhang}, \bibinfo{person}{Di
  Chen}, \bibinfo{person}{Hui Lu}, \bibinfo{person}{Wenlin Zhao},
  \bibinfo{person}{Lele Yu}, \bibinfo{person}{Xionghang Xie},
  \bibinfo{person}{Shiru Ren}, \bibinfo{person}{Xiang Sun},
  \bibinfo{person}{Yaocheng Tan}, \bibinfo{person}{Peng Xu},
  \bibinfo{person}{Yuchao Zheng}, {and} \bibinfo{person}{Di Wu}.}
  \bibinfo{year}{2025}\natexlab{}.
\newblock \showarticletitle{LONGER: Scaling Up Long Sequence Modeling in
  Industrial Recommenders}. In \bibinfo{booktitle}{\emph{Proceedings of the
  19th ACM Conference on Recommender Systems}}. \bibinfo{pages}{247--256}.
\newblock
\href{https://doi.org/10.1145/3705328.3748065}{doi:\nolinkurl{10.1145/3705328.3748065}}


\bibitem[Chang et~al\mbox{.}(2023a)]%
        {twin2023}
\bibfield{author}{\bibinfo{person}{Jianxin Chang}, \bibinfo{person}{Chenbin
  Zhang}, \bibinfo{person}{Zhiyi Fu}, \bibinfo{person}{Xiaoxue Zang},
  \bibinfo{person}{Lin Guan}, \bibinfo{person}{Jing Lu}, \bibinfo{person}{Yiqun
  Hui}, \bibinfo{person}{Dewei Leng}, \bibinfo{person}{Yanan Niu},
  \bibinfo{person}{Yang Song}, {and} \bibinfo{person}{Kun Gai}.}
  \bibinfo{year}{2023}\natexlab{a}.
\newblock \showarticletitle{TWIN: TWo-stage Interest Network for Lifelong User
  Behavior Modeling in {CTR} Prediction at Kuaishou}. In
  \bibinfo{booktitle}{\emph{Proceedings of the 29th ACM SIGKDD Conference on
  Knowledge Discovery and Data Mining}}. \bibinfo{pages}{3785--3794}.
\newblock
\href{https://doi.org/10.1145/3580305.3599922}{doi:\nolinkurl{10.1145/3580305.3599922}}


\bibitem[Chang et~al\mbox{.}(2023b)]%
        {chang2023pepnet}
\bibfield{author}{\bibinfo{person}{Jianxin Chang}, \bibinfo{person}{Chenbin
  Zhang}, \bibinfo{person}{Yiqun Hui}, \bibinfo{person}{Dewei Leng},
  \bibinfo{person}{Yanan Niu}, \bibinfo{person}{Yang Song}, {and}
  \bibinfo{person}{Kun Gai}.} \bibinfo{year}{2023}\natexlab{b}.
\newblock \showarticletitle{{PEPNet}: Parameter and Embedding Personalized
  Network for Infusing with Personalized Prior Information}. In
  \bibinfo{booktitle}{\emph{Proceedings of the 29th ACM SIGKDD Conference on
  Knowledge Discovery and Data Mining}}. \bibinfo{pages}{3795--3804}.
\newblock
\href{https://doi.org/10.1145/3580305.3599759}{doi:\nolinkurl{10.1145/3580305.3599759}}


\bibitem[Chen et~al\mbox{.}(2019)]%
        {chen2019bst}
\bibfield{author}{\bibinfo{person}{Qiwei Chen}, \bibinfo{person}{Huan Zhao},
  \bibinfo{person}{Wei Li}, \bibinfo{person}{Pipei Huang}, {and}
  \bibinfo{person}{Wenwu Ou}.} \bibinfo{year}{2019}\natexlab{}.
\newblock \showarticletitle{Behavior Sequence Transformer for E-commerce
  Recommendation in Alibaba}. In \bibinfo{booktitle}{\emph{Proceedings of the
  1st International Workshop on Deep Learning Practice for High-Dimensional
  Sparse Data}}.
\newblock
\href{https://doi.org/10.1145/3326937.3341261}{doi:\nolinkurl{10.1145/3326937.3341261}}


\bibitem[Cheng et~al\mbox{.}(2016)]%
        {cheng2016wide}
\bibfield{author}{\bibinfo{person}{Heng-Tze Cheng} {et~al\mbox{.}}}
  \bibinfo{year}{2016}\natexlab{}.
\newblock \showarticletitle{Wide \& Deep Learning for Recommender Systems}. In
  \bibinfo{booktitle}{\emph{Proceedings of the 1st Workshop on Deep Learning
  for Recommender Systems}}. \bibinfo{pages}{7--10}.
\newblock
\href{https://doi.org/10.1145/2988450.2988454}{doi:\nolinkurl{10.1145/2988450.2988454}}


\bibitem[Guan et~al\mbox{.}(2025)]%
        {guan2025stca}
\bibfield{author}{\bibinfo{person}{Lin Guan}, \bibinfo{person}{Jia-Qi Yang},
  \bibinfo{person}{Zhishan Zhao}, \bibinfo{person}{Beichuan Zhang},
  \bibinfo{person}{Bo Sun}, \bibinfo{person}{Xuanyuan Luo},
  \bibinfo{person}{Jinan Ni}, \bibinfo{person}{Xiaowen Li},
  \bibinfo{person}{Yuhang Qi}, \bibinfo{person}{Zhifang Fan},
  \bibinfo{person}{Hangyu Wang}, \bibinfo{person}{Qiwei Chen},
  \bibinfo{person}{Yi Cheng}, \bibinfo{person}{Feng Zhang}, {and}
  \bibinfo{person}{Xiao Yang}.} \bibinfo{year}{2025}\natexlab{}.
\newblock \bibinfo{title}{Make It Long, Keep It Fast: End-to-End 10k-Sequence
  Modeling at Billion Scale on Douyin}.
\newblock
\showeprint[arxiv]{2511.06077}~[cs.LG]


\bibitem[Gui et~al\mbox{.}(2023)]%
        {gui2023hiformer}
\bibfield{author}{\bibinfo{person}{Huan Gui}, \bibinfo{person}{Ruoxi Wang},
  \bibinfo{person}{Ke Yin}, \bibinfo{person}{Long Jin}, \bibinfo{person}{Maciej
  Kula}, \bibinfo{person}{Taibai Xu}, \bibinfo{person}{Lichan Hong}, {and}
  \bibinfo{person}{Ed~H. Chi}.} \bibinfo{year}{2023}\natexlab{}.
\newblock \bibinfo{title}{{HiFormer}: Heterogeneous Feature Interactions
  Learning with Transformers for Recommender Systems}.
\newblock
\showeprint[arxiv]{2311.05884}~[cs.IR]


\bibitem[Guo et~al\mbox{.}(2017)]%
        {guo2017deepfm}
\bibfield{author}{\bibinfo{person}{Huifeng Guo}, \bibinfo{person}{Ruiming
  Tang}, \bibinfo{person}{Yunming Ye}, \bibinfo{person}{Zhenguo Li}, {and}
  \bibinfo{person}{Xiuqiang He}.} \bibinfo{year}{2017}\natexlab{}.
\newblock \showarticletitle{{DeepFM}: A Factorization-Machine based Neural
  Network for {CTR} Prediction}. In \bibinfo{booktitle}{\emph{Proceedings of
  the 26th International Joint Conference on Artificial Intelligence}}.
  \bibinfo{pages}{1725--1731}.
\newblock


\bibitem[Huang et~al\mbox{.}(2019)]%
        {huang2019fibinet}
\bibfield{author}{\bibinfo{person}{Tongwen Huang}, \bibinfo{person}{Zhaozi
  Zhang}, {and} \bibinfo{person}{Junlin Zhang}.}
  \bibinfo{year}{2019}\natexlab{}.
\newblock \showarticletitle{{FiBiNET}: Combining Feature Importance and
  Bilinear Feature Interaction for Click-Through Rate Prediction}. In
  \bibinfo{booktitle}{\emph{Proceedings of the 13th ACM Conference on
  Recommender Systems}}. \bibinfo{pages}{169--177}.
\newblock
\href{https://doi.org/10.1145/3298689.3347033}{doi:\nolinkurl{10.1145/3298689.3347033}}


\bibitem[Huang et~al\mbox{.}(2026b)]%
        {mixformer2026}
\bibfield{author}{\bibinfo{person}{Xu Huang}, \bibinfo{person}{Hao Zhang},
  \bibinfo{person}{Zhifang Fan}, \bibinfo{person}{Yunwen Huang},
  \bibinfo{person}{Zhuoxing Wei}, \bibinfo{person}{Zheng Chai},
  \bibinfo{person}{Jinan Ni}, \bibinfo{person}{Yuchao Zheng}, {and}
  \bibinfo{person}{Qiwei Chen}.} \bibinfo{year}{2026}\natexlab{b}.
\newblock \bibinfo{title}{MixFormer: Co-Scaling Up Dense and Sequence in
  Industrial Recommenders}.
\newblock
\showeprint[arxiv]{2602.14110}~[cs.IR]


\bibitem[Huang et~al\mbox{.}(2026a)]%
        {hyformer2026}
\bibfield{author}{\bibinfo{person}{Yunwen Huang}, \bibinfo{person}{Shiyong
  Hong}, \bibinfo{person}{Xijun Xiao}, \bibinfo{person}{Jinqiu Jin},
  \bibinfo{person}{Xuanyuan Luo}, \bibinfo{person}{Zhe Wang},
  \bibinfo{person}{Zheng Chai}, \bibinfo{person}{Shikang Wu},
  \bibinfo{person}{Yuchao Zheng}, {and} \bibinfo{person}{Jingjian Lin}.}
  \bibinfo{year}{2026}\natexlab{a}.
\newblock \bibinfo{title}{HyFormer: Revisiting the Roles of Sequence Modeling
  and Feature Interaction in {CTR} Prediction}.
\newblock
\showeprint[arxiv]{2601.12681}~[cs.IR]


\bibitem[Kang and McAuley(2018)]%
        {kang2018sasrec}
\bibfield{author}{\bibinfo{person}{Wang-Cheng Kang} {and}
  \bibinfo{person}{Julian McAuley}.} \bibinfo{year}{2018}\natexlab{}.
\newblock \showarticletitle{Self-Attentive Sequential Recommendation}. In
  \bibinfo{booktitle}{\emph{Proceedings of the IEEE International Conference on
  Data Mining}}. \bibinfo{pages}{197--206}.
\newblock


\bibitem[Kaplan et~al\mbox{.}(2020)]%
        {kaplan2020scaling}
\bibfield{author}{\bibinfo{person}{Jared Kaplan}, \bibinfo{person}{Sam
  McCandlish}, \bibinfo{person}{Tom Henighan}, \bibinfo{person}{Tom~B. Brown},
  \bibinfo{person}{Benjamin Chess}, \bibinfo{person}{Rewon Child},
  \bibinfo{person}{Scott Gray}, \bibinfo{person}{Alec Radford},
  \bibinfo{person}{Jeffrey Wu}, {and} \bibinfo{person}{Dario Amodei}.}
  \bibinfo{year}{2020}\natexlab{}.
\newblock \bibinfo{title}{Scaling Laws for Neural Language Models}.
\newblock
\showeprint[arxiv]{2001.08361}~[cs.LG]


\bibitem[Li et~al\mbox{.}(2025)]%
        {infnet2025}
\bibfield{author}{\bibinfo{person}{Kaiyuan Li}, \bibinfo{person}{Dongdong Mao},
  \bibinfo{person}{Yongxiang Tang}, \bibinfo{person}{Yanhua Cheng},
  \bibinfo{person}{Yanxiang Zeng}, \bibinfo{person}{Chao Wang},
  \bibinfo{person}{Xialong Liu}, {and} \bibinfo{person}{Peng Jiang}.}
  \bibinfo{year}{2025}\natexlab{}.
\newblock \bibinfo{title}{{INFNet}: A Task-aware Information Flow Network for
  Large-Scale Recommendation Systems}.
\newblock
\showeprint[arxiv]{2508.11565}~[cs.IR]


\bibitem[Lian et~al\mbox{.}(2018)]%
        {lian2018xdeepfm}
\bibfield{author}{\bibinfo{person}{Jianxin Lian}, \bibinfo{person}{Xiaohuan
  Zhou}, \bibinfo{person}{Fuzheng Zhang}, \bibinfo{person}{Zhongxia Chen},
  \bibinfo{person}{Xing Xie}, {and} \bibinfo{person}{Guangzhong Sun}.}
  \bibinfo{year}{2018}\natexlab{}.
\newblock \showarticletitle{xDeepFM: Combining Explicit and Implicit Feature
  Interactions for Recommender Systems}. In
  \bibinfo{booktitle}{\emph{Proceedings of the 24th ACM SIGKDD International
  Conference on Knowledge Discovery \& Data Mining}}.
  \bibinfo{pages}{1754--1763}.
\newblock
\href{https://doi.org/10.1145/3219819.3220023}{doi:\nolinkurl{10.1145/3219819.3220023}}


\bibitem[Ma et~al\mbox{.}(2018)]%
        {ma2018mmoe}
\bibfield{author}{\bibinfo{person}{Jiaqi Ma}, \bibinfo{person}{Zhe Zhao},
  \bibinfo{person}{Xinyang Yi}, \bibinfo{person}{Jilin Chen},
  \bibinfo{person}{Lichan Hong}, {and} \bibinfo{person}{Ed~H. Chi}.}
  \bibinfo{year}{2018}\natexlab{}.
\newblock \showarticletitle{Modeling Task Relationships in Multi-Task Learning
  with Multi-Gate Mixture-of-Experts}. In \bibinfo{booktitle}{\emph{Proceedings
  of the 24th ACM SIGKDD International Conference on Knowledge Discovery \&
  Data Mining}}. \bibinfo{pages}{1930--1939}.
\newblock
\href{https://doi.org/10.1145/3219819.3220007}{doi:\nolinkurl{10.1145/3219819.3220007}}


\bibitem[Naumov et~al\mbox{.}(2019)]%
        {naumov2019dlrm}
\bibfield{author}{\bibinfo{person}{Maxim Naumov} {et~al\mbox{.}}}
  \bibinfo{year}{2019}\natexlab{}.
\newblock \bibinfo{title}{Deep Learning Recommendation Model for
  Personalization and Recommendation Systems}.
\newblock
\showeprint[arxiv]{1906.00091}~[cs.IR]


\bibitem[Perez et~al\mbox{.}(2018)]%
        {perez2018film}
\bibfield{author}{\bibinfo{person}{Ethan Perez}, \bibinfo{person}{Florian
  Strub}, \bibinfo{person}{Harm de Vries}, \bibinfo{person}{Vincent Dumoulin},
  {and} \bibinfo{person}{Aaron Courville}.} \bibinfo{year}{2018}\natexlab{}.
\newblock \showarticletitle{{FiLM}: Visual Reasoning with a General
  Conditioning Layer}. In \bibinfo{booktitle}{\emph{Proceedings of the AAAI
  Conference on Artificial Intelligence}}, Vol.~\bibinfo{volume}{32}.
\newblock


\bibitem[Pi et~al\mbox{.}(2020)]%
        {pi2020sim}
\bibfield{author}{\bibinfo{person}{Qi Pi}, \bibinfo{person}{Guorui Ge},
  \bibinfo{person}{Zhoujian Zhang}, \bibinfo{person}{Xiaoqiang Zhu},
  \bibinfo{person}{Han Zhu}, \bibinfo{person}{Kun Gai}, \bibinfo{person}{Ren
  Zhang}, {and} \bibinfo{person}{Xiao Han}.} \bibinfo{year}{2020}\natexlab{}.
\newblock \showarticletitle{Search-based User Interest Modeling with Lifelong
  Sequential Behavior Data for Click-Through Rate Prediction}. In
  \bibinfo{booktitle}{\emph{Proceedings of the 29th ACM International
  Conference on Information and Knowledge Management}}.
  \bibinfo{pages}{2685--2692}.
\newblock
\href{https://doi.org/10.1145/3340531.3344273}{doi:\nolinkurl{10.1145/3340531.3344273}}


\bibitem[Song et~al\mbox{.}(2019)]%
        {song2019autoint}
\bibfield{author}{\bibinfo{person}{Weiping Song}, \bibinfo{person}{Chence Shi},
  \bibinfo{person}{Zhiping Xiao}, \bibinfo{person}{Zhijian Duan},
  \bibinfo{person}{Yong Xu}, \bibinfo{person}{Ming Zhang}, {and}
  \bibinfo{person}{Jian Tang}.} \bibinfo{year}{2019}\natexlab{}.
\newblock \showarticletitle{AutoInt: Automatic Feature Interaction Learning via
  Self-Attentive Neural Networks}. In \bibinfo{booktitle}{\emph{Proceedings of
  the 28th ACM International Conference on Information and Knowledge
  Management}}. \bibinfo{pages}{1165--1174}.
\newblock
\href{https://doi.org/10.1145/3357384.3357906}{doi:\nolinkurl{10.1145/3357384.3357906}}


\bibitem[Tang et~al\mbox{.}(2020)]%
        {tang2020ple}
\bibfield{author}{\bibinfo{person}{Hongyu Tang}, \bibinfo{person}{Junning Liu},
  \bibinfo{person}{Ming Zhao}, {and} \bibinfo{person}{Xudong Gong}.}
  \bibinfo{year}{2020}\natexlab{}.
\newblock \showarticletitle{Progressive Layered Extraction ({PLE}): A Novel
  Multi-Task Learning ({MTL}) Model for Personalized Recommendations}. In
  \bibinfo{booktitle}{\emph{Proceedings of the 14th ACM Conference on
  Recommender Systems}}. \bibinfo{pages}{269--278}.
\newblock
\href{https://doi.org/10.1145/3383313.3412236}{doi:\nolinkurl{10.1145/3383313.3412236}}


\bibitem[Tolstikhin et~al\mbox{.}(2021)]%
        {tolstikhin2021mlpmixer}
\bibfield{author}{\bibinfo{person}{Ilya~O. Tolstikhin}, \bibinfo{person}{Neil
  Houlsby}, \bibinfo{person}{Alexander Kolesnikov}, \bibinfo{person}{Lucas
  Beyer}, \bibinfo{person}{Xiaohua Zhai}, \bibinfo{person}{Thomas Unterthiner},
  \bibinfo{person}{Jessica Yung}, \bibinfo{person}{Andreas Steiner},
  \bibinfo{person}{Daniel Keysers}, \bibinfo{person}{Jakob Uszkoreit},
  \bibinfo{person}{Mario Lucic}, {and} \bibinfo{person}{Alexey Dosovitskiy}.}
  \bibinfo{year}{2021}\natexlab{}.
\newblock \showarticletitle{{MLP-Mixer}: An All-{MLP} Architecture for Vision}.
  In \bibinfo{booktitle}{\emph{Advances in Neural Information Processing
  Systems}}, Vol.~\bibinfo{volume}{34}. \bibinfo{pages}{24261--24272}.
\newblock


\bibitem[Vaswani et~al\mbox{.}(2017)]%
        {vaswani2017attention}
\bibfield{author}{\bibinfo{person}{Ashish Vaswani}, \bibinfo{person}{Noam
  Shazeer}, \bibinfo{person}{Niki Parmar}, \bibinfo{person}{Jakob Uszkoreit},
  \bibinfo{person}{Llion Jones}, \bibinfo{person}{Aidan~N. Gomez},
  \bibinfo{person}{{\L}ukasz Kaiser}, {and} \bibinfo{person}{Illia
  Polosukhin}.} \bibinfo{year}{2017}\natexlab{}.
\newblock \showarticletitle{Attention is All You Need}. In
  \bibinfo{booktitle}{\emph{Advances in Neural Information Processing
  Systems}}, Vol.~\bibinfo{volume}{30}.
\newblock


\bibitem[Wang et~al\mbox{.}(2017)]%
        {wang2017dcn}
\bibfield{author}{\bibinfo{person}{Ruoxi Wang}, \bibinfo{person}{Bin Fu},
  \bibinfo{person}{Gang Fu}, {and} \bibinfo{person}{Mingliang Wang}.}
  \bibinfo{year}{2017}\natexlab{}.
\newblock \showarticletitle{Deep \& Cross Network for Ad Click Predictions}. In
  \bibinfo{booktitle}{\emph{Proceedings of the ADKDD'17}}. \bibinfo{pages}{12}.
\newblock
\href{https://doi.org/10.1145/3124749.3124754}{doi:\nolinkurl{10.1145/3124749.3124754}}


\bibitem[Wang et~al\mbox{.}(2021)]%
        {wang2021dcnv2}
\bibfield{author}{\bibinfo{person}{Ruoxi Wang}, \bibinfo{person}{Rakesh
  Shivanna}, \bibinfo{person}{Derek Cheng}, \bibinfo{person}{Sagar Jain},
  \bibinfo{person}{Dandelion Lin}, \bibinfo{person}{Lichan Hong}, {and}
  \bibinfo{person}{Ed Chi}.} \bibinfo{year}{2021}\natexlab{}.
\newblock \showarticletitle{{DCN} {V2}: Improved Deep \& Cross Network and
  Practical Lessons for Web-scale Learning to Rank Systems}. In
  \bibinfo{booktitle}{\emph{Proceedings of the Web Conference 2021}}.
  \bibinfo{pages}{1785--1797}.
\newblock
\href{https://doi.org/10.1145/3442381.3450078}{doi:\nolinkurl{10.1145/3442381.3450078}}


\bibitem[Xi et~al\mbox{.}(2021)]%
        {xi2021aitm}
\bibfield{author}{\bibinfo{person}{Dongbo Xi}, \bibinfo{person}{Zhen Chen},
  \bibinfo{person}{Peng Yan}, \bibinfo{person}{Yinger Zhang},
  \bibinfo{person}{Yongchun Zhu}, \bibinfo{person}{Fuzhen Zhuang}, {and}
  \bibinfo{person}{Yu Chen}.} \bibinfo{year}{2021}\natexlab{}.
\newblock \showarticletitle{Modeling the Sequential Dependence among Audience
  Multi-step Conversions with Multi-task Learning in Targeted Display
  Advertising}. In \bibinfo{booktitle}{\emph{Proceedings of the 27th ACM SIGKDD
  International Conference on Knowledge Discovery \& Data Mining}}.
  \bibinfo{pages}{3745--3755}.
\newblock
\href{https://doi.org/10.1145/3447548.3467071}{doi:\nolinkurl{10.1145/3447548.3467071}}


\bibitem[Zeng et~al\mbox{.}(2024)]%
        {zeng2024interformer}
\bibfield{author}{\bibinfo{person}{Zhichen Zeng}, \bibinfo{person}{Xiaolong
  Liu}, \bibinfo{person}{Mengyue Hang}, \bibinfo{person}{Xiaoyi Liu},
  \bibinfo{person}{Qinghai Zhou}, \bibinfo{person}{Chaofei Yang},
  \bibinfo{person}{Yiqun Liu}, \bibinfo{person}{Yichen Ruan},
  \bibinfo{person}{Laming Chen}, \bibinfo{person}{Yuxin Chen}, {et~al\mbox{.}}}
  \bibinfo{year}{2024}\natexlab{}.
\newblock \bibinfo{title}{{InterFormer}: Towards Effective Heterogeneous
  Interaction Learning for Click-Through Rate Prediction}.
\newblock
\showeprint[arxiv]{2411.09852}~[cs.IR]


\bibitem[Zhai et~al\mbox{.}(2024)]%
        {zhai2024hstu}
\bibfield{author}{\bibinfo{person}{Jiahui Zhai}, \bibinfo{person}{Jiayu Liao},
  \bibinfo{person}{Liang Zhao}, \bibinfo{person}{Cheng Lin},
  \bibinfo{person}{Xinxing Wang}, \bibinfo{person}{Zhen Xu},
  \bibinfo{person}{Kun Zhang}, \bibinfo{person}{Min-Jae Kim},
  \bibinfo{person}{Chengxing Yu}, \bibinfo{person}{Weirui Zhao},
  \bibinfo{person}{Priya Jasin}, \bibinfo{person}{Yinfei Hu},
  \bibinfo{person}{Alan Shapiro}, \bibinfo{person}{Weijie Yu},
  \bibinfo{person}{Minmin Zheng}, \bibinfo{person}{Yiqiao Guan},
  \bibinfo{person}{James Li}, \bibinfo{person}{Xiaotong Ju},
  \bibinfo{person}{Yuyan Gong}, \bibinfo{person}{Jing Liu},
  \bibinfo{person}{Shuang Zhang}, \bibinfo{person}{Zhao Ma},
  \bibinfo{person}{Yuxin Zhao}, \bibinfo{person}{Lei Pao},
  \bibinfo{person}{Hui~Li Liang}, \bibinfo{person}{Susheel Kharbanda},
  \bibinfo{person}{Pratik Jain}, \bibinfo{person}{Hongyi Jiang},
  \bibinfo{person}{Aihan Elibol}, \bibinfo{person}{Lucas Klobuchar},
  \bibinfo{person}{Kent Langman}, \bibinfo{person}{Mehmet~fethi Erensoy},
  \bibinfo{person}{Muralidhar Shiraishi}, \bibinfo{person}{Da-Cheng Lee},
  \bibinfo{person}{Ellie Deng}, \bibinfo{person}{Jihoon Cho},
  \bibinfo{person}{Nam Yu}, \bibinfo{person}{Yuzheng Liu},
  \bibinfo{person}{Zhengyu Zhao}, \bibinfo{person}{Sumit Bhatnagar},
  \bibinfo{person}{Xiaolong Liu}, \bibinfo{person}{Yuekai Noghabaei},
  \bibinfo{person}{Karthik Parthasarathy}, \bibinfo{person}{Otto Lang},
  \bibinfo{person}{Aditya Krishnan}, \bibinfo{person}{Han Nguyen},
  \bibinfo{person}{Wei Chen}, \bibinfo{person}{Chuan Xu}, \bibinfo{person}{Yong
  Jia}, {and} \bibinfo{person}{Lijuan Wang}.} \bibinfo{year}{2024}\natexlab{}.
\newblock \showarticletitle{Actions Speak Louder than Words: Trillion-Parameter
  Sequential Transducers for Generative Recommendations}. In
  \bibinfo{booktitle}{\emph{Proceedings of the 41st International Conference on
  Machine Learning}} \emph{(\bibinfo{series}{Proceedings of Machine Learning
  Research}, Vol.~\bibinfo{volume}{235})}.
\newblock


\bibitem[Zhang et~al\mbox{.}(2024)]%
        {zhang2024wukong}
\bibfield{author}{\bibinfo{person}{Qi Zhang}, \bibinfo{person}{Kush Bhatia},
  \bibinfo{person}{Aviral Gupta}, \bibinfo{person}{Apurv Yadav},
  \bibinfo{person}{Adnan Li}, \bibinfo{person}{Zhiyi Zhang},
  \bibinfo{person}{Youlong Bao}, \bibinfo{person}{Borui Chen},
  \bibinfo{person}{Nikita Trushkin}, \bibinfo{person}{Mohammadhossein Bateni},
  \bibinfo{person}{Peiran Han}, \bibinfo{person}{Rathin Singh},
  \bibinfo{person}{Tong Bui}, \bibinfo{person}{Keigo Kojima},
  \bibinfo{person}{Nikolay Zhelev}, \bibinfo{person}{James Li},
  \bibinfo{person}{Jiayi He}, \bibinfo{person}{Iesh Sharan},
  \bibinfo{person}{Srinivas Gopalakrishnan}, \bibinfo{person}{Bing Yang},
  \bibinfo{person}{Paul Covington}, {and} \bibinfo{person}{Ed~H. Chi}.}
  \bibinfo{year}{2024}\natexlab{}.
\newblock \showarticletitle{Wukong: Towards a Scaling Law for Large-Scale
  Recommendation}. In \bibinfo{booktitle}{\emph{Proceedings of the 41st
  International Conference on Machine Learning}}
  \emph{(\bibinfo{series}{Proceedings of Machine Learning Research},
  Vol.~\bibinfo{volume}{235})}.
\newblock


\bibitem[Zhang et~al\mbox{.}(2025)]%
        {onetrans2025}
\bibfield{author}{\bibinfo{person}{Zhaoqi Zhang}, \bibinfo{person}{Haolei Pei},
  \bibinfo{person}{Jun Guo}, \bibinfo{person}{Tianyu Wang},
  \bibinfo{person}{Yufei Feng}, \bibinfo{person}{Hui Sun},
  \bibinfo{person}{Shaowei Liu}, {and} \bibinfo{person}{Aixin Sun}.}
  \bibinfo{year}{2025}\natexlab{}.
\newblock \bibinfo{title}{OneTrans: Unified Feature Interaction and Sequence
  Modeling with One Transformer in Industrial Recommender}.
\newblock
\showeprint[arxiv]{2510.26104}~[cs.IR]


\bibitem[Zhou et~al\mbox{.}(2019)]%
        {zhou2019dien}
\bibfield{author}{\bibinfo{person}{Guorui Zhou} {et~al\mbox{.}}}
  \bibinfo{year}{2019}\natexlab{}.
\newblock \showarticletitle{Deep Interest Evolution Network for Click-Through
  Rate Prediction}. In \bibinfo{booktitle}{\emph{Proceedings of the AAAI
  Conference on Artificial Intelligence}}, Vol.~\bibinfo{volume}{33}.
  \bibinfo{pages}{5941--5948}.
\newblock


\bibitem[Zhou et~al\mbox{.}(2018)]%
        {zhou2018din}
\bibfield{author}{\bibinfo{person}{Guorui Zhou}, \bibinfo{person}{Xiaoqiang
  Zhu}, \bibinfo{person}{Chenru Song}, \bibinfo{person}{Ying Fan},
  \bibinfo{person}{Han Zhu}, \bibinfo{person}{Xiao Ma}, \bibinfo{person}{Yang
  Tang}, \bibinfo{person}{Hui Jin}, \bibinfo{person}{Kun Li}, {and}
  \bibinfo{person}{Kun Gai}.} \bibinfo{year}{2018}\natexlab{}.
\newblock \showarticletitle{Deep Interest Network for Click-Through Rate
  Prediction}. In \bibinfo{booktitle}{\emph{Proceedings of the 24th ACM SIGKDD
  International Conference on Knowledge Discovery \& Data Mining}}.
  \bibinfo{pages}{1059--1068}.
\newblock
\href{https://doi.org/10.1145/3219819.3219823}{doi:\nolinkurl{10.1145/3219819.3219823}}


\bibitem[Zhu et~al\mbox{.}(2025)]%
        {rankmixer2025}
\bibfield{author}{\bibinfo{person}{Jie Zhu}, \bibinfo{person}{Zhifang Fan},
  \bibinfo{person}{Xiaoxie Zhu}, \bibinfo{person}{Yuchen Jiang},
  \bibinfo{person}{Hangyu Wang}, \bibinfo{person}{Xintian Han},
  \bibinfo{person}{Haoran Ding}, \bibinfo{person}{Xinmin Wang},
  \bibinfo{person}{Wenlin Zhao}, \bibinfo{person}{Zhen Gong},
  \bibinfo{person}{Huizhi Yang}, \bibinfo{person}{Zheng Chai},
  \bibinfo{person}{Zhe Chen}, \bibinfo{person}{Yuchao Zheng},
  \bibinfo{person}{Qiwei Chen}, \bibinfo{person}{Feng Zhang},
  \bibinfo{person}{Xun Zhou}, \bibinfo{person}{Peng Xu}, \bibinfo{person}{Xiao
  Yang}, \bibinfo{person}{Di Wu}, {and} \bibinfo{person}{Zuotao Liu}.}
  \bibinfo{year}{2025}\natexlab{}.
\newblock \showarticletitle{RankMixer: Scaling Up Ranking Models in Industrial
  Recommenders}. In \bibinfo{booktitle}{\emph{Proceedings of the 34th ACM
  International Conference on Information and Knowledge Management}}.
\newblock
\href{https://doi.org/10.1145/3746252.3761507}{doi:\nolinkurl{10.1145/3746252.3761507}}


\end{thebibliography}

\end{document}